# Contactless haptic display through magnetic field control

Xiong Lu, *Member, IEEE,* Yuxing Yan, Beibei Qi, Huang Qian, Junbin Sun, and Aaron Quigley

*Abstract*—Haptic rendering enables people to touch, perceive, and manipulate virtual objects in a virtual environment. Using six cascaded identical hollow disk electromagnets and a small permanent magnet attached to an operator's finger, this paper proposes and develops an untethered haptic interface through magnetic field control. The concentric hole inside the six cascaded electromagnets provides the workspace, where the 3D position of the permanent magnet is tracked with a Microsoft Kinect sensor. The driving currents of six cascaded electromagnets are calculated in real-time for generating the desired magnetic force. Offline data from an FEA (finite element analysis) based simulation, determines the relationship between the magnetic force, the driving currents, and the position of the permanent magnet. A set of experiments including the virtual object recognition experiment, the virtual surface identification experiment, and the user perception evaluation experiment were conducted to demonstrate the proposed system, where Microsoft HoloLens holographic glasses are used for visual rendering. The proposed magnetic haptic display leads to an untethered and non-contact interface for natural haptic rendering applications, which overcomes the constraints of mechanical linkages in tool-based traditional haptic devices.

## I. INTRODUCTION

HAPTICS is a general term used to describe something relating to or based on the sense of touch. Among all senses, the haptic system provides a unique and bi-directional communication channel for humans during their interactions with the real world around them. However, most current human-computer interaction (HCI) systems have focused primarily on visual rendering and auditory rendering. By contrast, haptic rendering provides a new way for human beings to actively perceive, touch, and explore objects in virtual environments (VEs) or remote environments. This has the potential to increase the quality of interaction, improve the fidelity of virtual reality, and enhance our level of understanding of complex data sets. As a result, haptic rendering has become an emerging research topic in HCI and has been effectively used in various applications, such as medical training [1], [2], mechanical manufacturing [3], aerospace [4], and education [5], [6], teleoperation [7], [8], human-robot collaboration [9], and virtual interpersonal touch [10]. However, the means by which we can experience such haptic outputs or provide such inputs remains largely contact-based.

In a haptic system, the "haptic interface" provides a bidirectional link between the human operator and the VE, which allows the operator to interact with virtual objects by providing them with force/torque. To date, different types of haptic interfaces have been developed. And several haptic devices have even been commercialized, such as Geomagic Touch series (formerly Sensable PHANTOM) from 3D System, Omega, Delta and Sigma haptic devices from Force Dimension, and the High Definition Haptic Device from Quanser. Such sustained commercial and research interest has driven widening attention to haptic interfaces in recent years.

However, current haptic devices typically use mechanical linkages or exoskeleton structures to impart forces and torques to the operator. The former type is limited to tool-based haptic interactions (such as a stylus) or what might be broadly considered "instrumental interactions [11]". These are not always the intuitive, or indeed natural ways, for humans to interact with the world. Moreover, the mechanical linkage can be a constraint to humans, i.e., restricting free mobility. While the exoskeleton structures can be an uncomfortable burden for humans during their operations. Providing force/torque to a person without complex contact-based assemblage is challenging.

The objective of this paper is to achieve a contactless magnetic haptic interface, by means of a controllable electromagnetic field and a small permanent magnet attached to the human operator's fingertip. This allows for the generation of magnetic forces while also allowing the human hand to move freely. To achieve this, the controllable electromagnetic field is actuated by six cascaded identical circular disk electromagnets. Concentric holes inside each disk electromagnets afford the operator a workspace in the system and a Microsoft Kinect is used for position tracking of the permanent magnet on the operator's fingertip.

The rest of this paper is organized as follows. Related work is presented in Section II. Section III provides design details of the proposed contactless one-dimensional (1D) haptic system. The magnetic field distribution in the electromagnetic coil space and the force analysis of the permanent magnet based on finite-element-method (FEM) is analyzed in Section IV.

This work was supported in part by the National Natural Science Foundation of China under Grant No. 61773205, by the Natural Science Foundation of Jiangsu Province under Grant No. BK20211186, by the scholarship from China Scholarship Council (CSC) under No. 201906835020, and by the Fundamental Research Funds for the Central Universities (Nanjing University of Aeronautics and Astronautics) under No. NS2019018. *(Corresponding author: Xiong Lu)*

Xiong Lu, Yuxing Yan,Beibei Qi, Huang Qian, and Junbin Sun are with the College of Automation Engineering, Nanjing University of Aeronautics and Astronautics, Nanjing 211106, China (email: luxiong@nuaa.edu.cn; yanyuxing@nuaa.edu.cn; qibeibei1010@126.com; 15150651110@163.com).

Aaron Quigley is currently a Professor in the School of Computer Science and Engineering, UNSW, in Sydney Australia and head of school (e-mail: a.quigley@unsw.edu.au).

Section V illustrates recognition experiments of virtual objects and experimental results. While Section VI concludes and provides research directions for future work.

## II. RELATED WORK

Conventional haptic rendering methods utilize various mechanical assemblages to simulate the haptic feedback of virtual objects and use mechanical linkages to impart forces to people.

Craig Carignan et al. [12] proposed a 6DOF exoskeleton device for shoulder joint rehabilitation training, MGA(Maryland-Georgetown-Army), which provides force feedback through force/torque sensors. However, due to the mechanical linkages, there is a certain viscous resistance during work, which reduces the reality of haptic rendering. E. C. Lovasz et al. [13] developed an elbow exoskeleton haptic feedback device for specific needs in space applications, which can control the movement of the robot where a user feels the corresponding haptic feedback. T. L. Baldi et al. [14] designed a wearable glove GESTO based on inertial sensors and magnetic sensors for hand tracking, which are utilized to enhance hand perception and touch virtual objects in virtual game scenarios. In addition, joystick-type haptic rendering devices have been proposed, such as the 6DOF multi-finger haptic interface developed by H. H. Qin et al [15], and the novel haptic device VirSense designed by A. Mashayekhi et al [16]. Although wearable and joystick haptic rendering devices have the advantages of high control accuracy and large output force, their development are restricted by the mechanical assemblages, special system structures and high costs to varying extents.

To overcome the shortcomings of the traditional haptic rendering devices, some non-contact haptic rendering devices based on ultrasonic waves [17], air jets [18], and magnetic forces [19] have been proposed and developed. Such non-contact haptic rendering devices typically eliminate the influence of friction and inertia, increasing the sense of reality and immersion in haptic rendering systems.

Recent work has focused on providing rich, yet contact free haptic feedback, overcoming the need for expensive and complex robotic-arm like elements [20]. Compared with other types of forces, the electromagnetic force is widely used for designing non-contact haptic rendering devices. The magnetic field generated by the coils can directly generate forces on the permanent magnet, on the tip of the stylus or the human fingertip, there will not be additional mechanical devices or moving parts. Haptic interface devices based on Lorentz force magnetic levitation [21], [22], [23], [24], [25] can provide motion and force feedback in 6-DoF with high closed-loop control bandwidth and position resolution. A micro-positioning stage in 6-DoF based on magnetic levitation was presented in [26] for a haptic teleoperated control system. To augment the tissue stiffness perception in the virtual environment, Tong et al. designed a novel magnetic levitation haptic device and proposed an algorithm to calculate the best coil attitude [27]. They verified that the proposed haptic device showed high accuracy (0.28 mm) of real-time tracking for the magnetic stylus and the low power consumption of the coil configuration. Hollis R L. et al. [28] designed the butterfly magnetic levitation device that eliminates the bulky links, cables, and general mechanical complexity of other haptic devices, by using a lightweight moving part floating on a magnetic field. However, the operating space of this device is limited, with a horizontal range of 25mm and a rotation angle of 20°.

The disadvantage of existing Lorentz levitation devices is their available workspaces are limited to 50mm or less, which limits their usage in general HCI and hence haptic interaction applications. A possible solution to this problem is to generate a magnetic field in the air from the electromagnets array, where the force field can be felt as the magnetic repulsive force. A 15-by-15 electromagnet array driven by the direct current was designed to generate the magnetic field in [29], where users can feel volumetric shapes with little attachment on the hand by applying magnetic repelling force. This system was modeled and tested via FEM simulations, which can result in a smooth force field and accurate force exertion on the control points. Adel et al. [30], [31] developed a position-based impedance electromagnetic haptic interface based on 9 and 25 coil arrays, respectively. The system, with a workspace of 150mm * 150mm * 20mm, can generate 50mN electromagnetic force, which can be used for subjects to distinguish the general contour of 3D virtual objects. P. Berkelman et al. [32] proposed a novel haptic device using a thin flat array of coils and sensors to generate force feedback on a magnet held near the display screen. The active area of this system is approximately 120mm * 120 mm, and 3D forces can be produced in any direction, both parallel and normal to the screen surface. A. Abler et al. [33] explored a new method for passive magnet interaction in a non-planar array and designed a spherical handheld controller that provides up to 200 mN of haptic force to the user's palm through a passive magnetic needle.

To increase the electromagnetic force, some haptic interaction devices based on large coils have been explored. An electromagnetic force feedback system with a pair of box-shaped coils and an array of magnetometer sensors was proposed in [34]. This system, with a workspace of 100 mm * 100 mm, delivers planar force feedback to a magnet, fixed to a user stylus or the human fingertip, to enable haptic force interaction with virtual objects and fixtures modeled in software. Nevertheless, only planar forces were generated in the system. J. Petruska and J. Abbott [35] proposed the Omnimagnet composed of three electromagnets, widely used in remote manipulation [36]. By controlling the current in each of the three coils, a magnetic dipole moment can be created in any direction of any desired strength. Also, the developed haptic virtual fixtures helped a human-machine collaborative system perform a task by limiting movement into restricted regions and/or influencing movement along desired paths [37]. A. Adel et al. also focused on magnetic localization for an electromagnetic-based haptic interface, using two identical arrays of three-dimensional magnetic field sensors to estimate the position of the magnetic dipole [38]. J. J. Zarate et al. [39] proposed a novel contact-free volumetric haptic feedback device, which is composed of a spherical electromagnet with a diameter of 60 mm that can generate a force of up to 1N on a directional tool embedded with a permanent magnet. Based on a 3D layer-by-layer constructed omnidirectional electromagnet (65 mm outer diameter with a 30 mm diameter soft iron core inside), a small permanent magnet, and a hall sensing array, T. Langerak et al. presented a self-contained 3D haptic system [40]. R. Zhang et al. [41] explored the mapping of 6D vibrotactile stimulus rendered at the haptic interaction point (HIP) of a kinesthetic haptic interface to an equivalent 1D stimulus when the stylus is accurately held. Generally, design and performance issues involved with untethered magnetic haptic interfaces can refer to the research conclusions of [20]. Furthermore, magnetic haptic interfaces are also designed for minimally invasive surgery [42], teleoperation [43], and micromanipulation platforms [44].

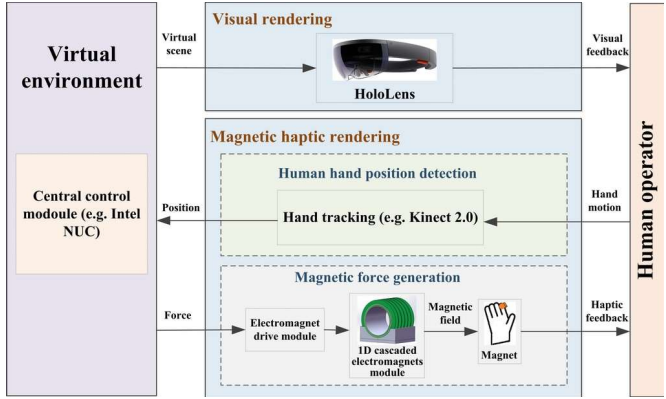

Fig. 1. The structure diagram of the magnetic force actuated 1D haptic system.

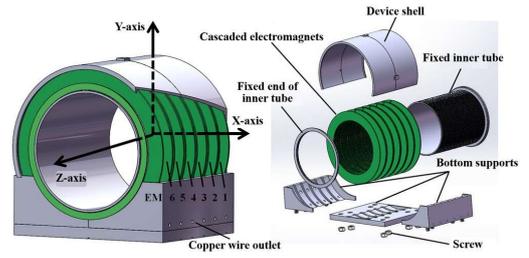

Fig. 2. The design diagram of cascaded electromagnets module (Left: assembly view, right: exploded view.)

In this paper, a contactless haptic interface, using six cascaded identical hollow disk electromagnets and a small finger wearable permanent magnet, has been designed. Compared with the height-limited volumetric workspaces above the array plane for electromagnet array-based haptic devices [31], [34], [33], [29], and the volumetric workspace surrounding the box coils [35], [36] or sphere coils [40], [39], [41] for large-coil based haptic devices, our haptic interface has a relatively large cylindrical workspace of Ø 210mm * H 234mm inside the six identical hollow disk electromagnets, which importantly can accommodate the entire hand of a human operator and enables dexterous interaction operations. Moreover, the cascaded-electromagnet structure in our proposed haptic interface enables the independent current driving for each identical hollow disk electromagnet, leading to a more flexible magnetic force control and regulation.

## III. SYSTEM DESIGN

The architecture of a typical haptic system based on the proposed 1D magnetic haptic device is shown in Fig. 1. The system is composed of the visual rendering module and the magnetic haptic rendering module. The haptic rendering module can be further divided into the magnetic force generation module and human hand position tracking module, i.e. the position of the permanent magnet.

### A. Magnetic force generation module

In our proposed method, the electromagnetic principle is used to generate interacting forces in a contactless style. To generate a strong controllable magnetic field in a relatively large workspace, an electromagnetic module composed of 6 cascaded identical disk-like solenoid coils (Fig. 2) is designed. The optimal solution for selecting the coil parameters is a trade-off between magnetic force and workspace. To generate required electromagnetic forces and guarantee a large enough workspace for a human's hand of average size, magnetic force analyses with the FEA-based method have been carried out for selecting these parameters of our final optimal design of coils. Each coil was constructed by two strands of enameled copper wire with a diameter of 0.67 mm to reduce the required driving voltage, being within 80V range of the KEYSIGHT N5768A DC power module. For each coil, the 3D printed frame with an inner diameter of 220mm and an axial thickness of 35mm is used to mount the copper wire, leading to about 3000 turns with a resistance of 17.6 Ohm. The concentric holes inside 6 coils form a cylindrical workspace of $\varphi$ 210mm * $H$ 234mm, which is suitable for the motion of an adult's single hand with an average size. The device shell with permalloy material in Fig. 2 is utilized for magnetic shielding. A 3D printed support base with several parts is also designed to support the weight of 6 coils. The detailed parameters for each coil are shown in Table I.

Two SA306AHU chips (supply range -8.5V to 60V; output current per phase 8A continuous) are used to regulate the driving current and change its direction for each coil with the pulse width modulation (PWM) signals, which are generated by a Cortex-M4 Microcontroller based embedded system module (EK-TM4C1294XL from TI). A total of 12 sets of PWM signals have been used for driving six coils and generating a one-dimensional controllable electromagnetic field.

TABLE I
PARAMETERS OF A SINGLE ELECTROMAGNET COIL

| Inner radius (mm) | Outer radius (mm) | Thickness (mm) | Diameter of copper wire (mm) | Number of coil turns (turn) |
|---|---|---|---|---|
| 115.0 | 145.0 | 30.0 | 0.67 | 1500 |

A disk-shaped permanent magnet attached to the human operator's fingertip is used to render the forces as the human

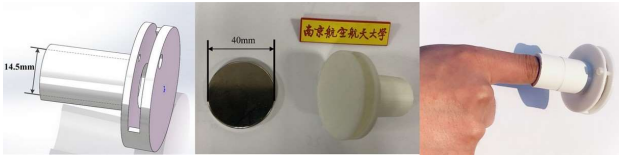

Fig. 3. The 3D-printed finger wearable module and the permanent magnet for mounting on the operator's fingertip. (Left: 3D design, Middle: Photos, Right: Mounted on fingertip)

operator interacts with virtual objects. The concentric hole in each coil forms the workspace for haptic rendering. According to the FEM (Finite element method) based simulation results for maximum magnetic forces, an off-the-shelf Neodymium-iron-boron (N35) permanent magnet with the dimension of $\varnothing$ 40mm * $H$ 4mm is chosen. This dimensional size of the permanent magnet is suitable for wearing on the operator's index finger of average adult size, by mounting inside a 3D-printed finger wearable module (Fig. 3). The inner diameter of the finger wearable module is set to 14.5 mm to meet ergonomic requirements.

### B. Finger position tracking module

In our current proposed system, the permanent magnet will move with the operator's finger in a steady posture. The position of the operator's fingertip is measured in real-time for contact detection and force calculation in haptic rendering. A Microsoft Kinect 2.0 sensor, with high measurement accuracy and real-time performance [45], is used for tracking the position of the permanent magnet.

An Intel NUC6i7KYK has been chosen to be the central control and computing platform for implementing the Kinect based finger position tracking algorithms, as well as collision detecting, force calculating, and the driving currents calculating of 6 coils for generating the desired forces. In each rendering loop, the driving currents for 6 coils in the form of 6 duty cycle values are sent to the electromagnet control module through an RS-232 serial interface.

### C. Visual rendering module

A Microsoft HoloLens has been used for rendering a mixed reality visual scene to the human operator, including both real and holographic content. The HoloLens scene-keeping function can be realized through the programming of the World Anchor script, which provides a method to keep virtual objects in specific locations and states. This function guarantees the stability of the holographic objects. Moreover, the HoloLens Wi-Fi module is used to transfer bidirectional data between the HoloLens and the NUC central control module.

## IV. METHOD FOR ELECTROMAGNETIC FORCE ACTUATION

In the proposed system, the most challenging work will be revealing the relationship between the generated magnetic force on a permanent magnet, the driving currents of six cascaded coil electromagnets, and the position of the permanent magnet in the workspace. Although several research explorations have been carried out for calculating the axial force between a coil and a cylindrical permanent magnet [46] or magnetic levitation force [47], the challenge in our system is more complicated. Our strategy for solving this problem is, (1) Simulating the relationship between magnetic forces and the driving currents in grid locations in the 3D workspace, (2) Generating offline data representing the above relationship, (3) Calculating the driving currents of 6 cascaded coils in real-time for any desired forces with interpolation method based on the offline data.

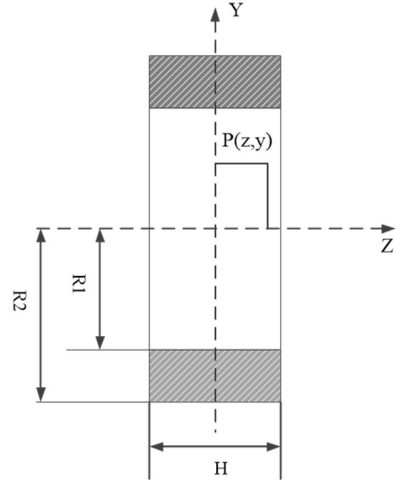

Fig. 4. Diagram of a single-coil electromagnet

As shown in Fig. 4, the central axis of the one single-coil electromagnet is defined as the Z-axis. The black shadow is the part where the coil is wound. $R1$ is the inner diameter of the solenoid while $R2$ is the outer diameter, and $H$ is the length of the solenoid along the axis. If the current of the solenoid and the number of winding turns are given, the magnetic flux density at any point $P$ in the space can be obtained.

In this paper, we use finite element method (ANSYS Student) to compute the relationships (field values) at a grid of points to analyze the distribution of electromagnetic field and the electromagnetic force exerted on the permanent magnet. Considering the practical design, actuation and implementation costs, six disc-like solenoid coils are proposed in our system for generating a controllable magnetic field. The six coils of the cascaded electromagnets module have the same size and the same number of turns, where the current magnitude and direction of each coil can be controlled with SA306 based drive modules separately.

In the following simulation research, a single solenoid coil is used to analyze the relationship between the magnetic force, driving current, and permanent magnet's position. Then, the simulation analysis of 6 solenoid coils is obtained with the magnetic field superposition principle. Finally, a cascade electromagnet control method based on offline simulation data was proposed.

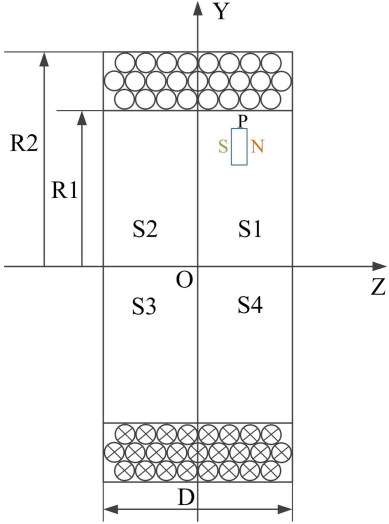

Fig. 5. The operating workspace of a single-coil electromagnet.

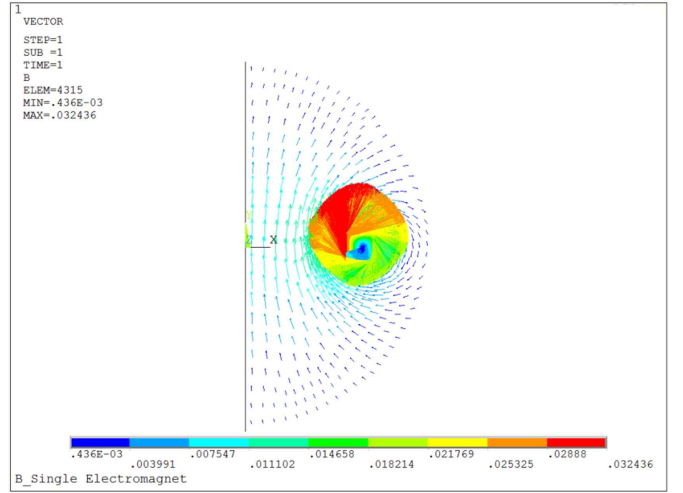

Fig. 6. The vector distribution diagram of the electromagnetic field.

## A. Simulation analysis of cascaded electromagnets module

Two parameters, i.e., the distribution of magnetic flux density in the single-coil electromagnet and the forces between the permanent magnet and the coil, have been first studied. The schematic diagram of the single-coil electromagnet is shown in Fig. 5 (the single-coil electromagnet with inner radius $R1$ of 115mm, outer radius $R2$ of 145mm and thickness $D$ of 30mm). The ranges [-215mm, 215mm] in the Z-axis and [-110mm, 110mm] in the Y-axis are chosen for the simulation. The entire workspace of interest is divided into four parts: $S1$, $S2$, $S3$, and $S4$. From the symmetry of the electromagnetic principle, it can be known that the electromagnetic force on the permanent magnet in $S1$ is the same as that of $S4$, and the force with the same magnitude and opposite direction exists in the area of $S2$ and $S3$. Therefore, the FEA-based magnetic force analysis in $S1$ has been carried out. Detailed parameters of the single-coil electromagnet are shown in table I.

The polyester enameled copper wire with a diameter of 0.67mm and the maximum safe current of 1.6 A is chosen to construct our electromagnet coils. The direction of the driving current is shown in Fig. 5. When the excitation current is set to 1.6A, the vector distribution diagram of the electromagnetic field of a single electromagnet is shown in Fig. 6. Then, the magnitude distribution of the magnetic flux density within the operating spaces $S1$ and $S2$ of a single-coil electromagnet can be obtained (Fig. 7). It shows that the magnetic flux density increases along the positive direction of the Z-axis and then decreases when the Y-axis coordinate remains unchanged. The magnetic flux density reaches the maximum value at the position where Z- and Y-axis coordinate are 0 mm and 110 mm respectively.

Fig. 8 illustrate the force exerted on the permanent magnet (N35 material with a magnetic coercivity of 8.59 * 105 A/m, Ø 40mm*$H$ 4mm) when the driving current of the single-coil electromagnet is 1.6 A. The coordinate symbol ($P\_X$, $P\_Y$, $P\_Z$) is used for denoting the position of permanent magnet in the workspace. Considering the volume of the permanent magnet, its central point position remains in the range of [0, 90mm] in the Y-axis for area $S1$ during the simulation. The grid step for the permanent magnet's position is 5 mm, that is, the workspace for the permanent magnet in the simulation is divided into a grid of 5 mm * 5 mm on the $S1$.

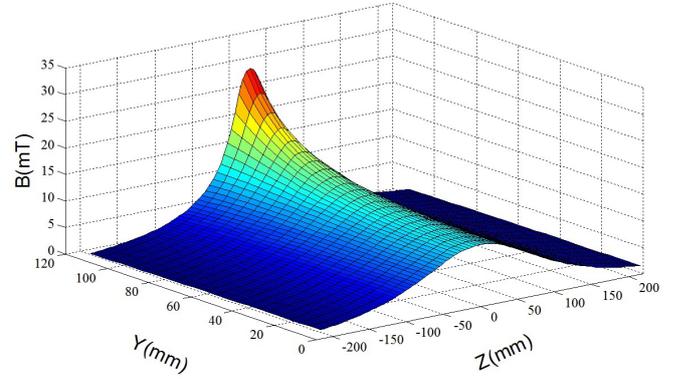

Fig. 7. Simulation of magnetic flux density (I=1.6A)

It can be seen that when a current is applied to the single-coil electromagnet, the electromagnetic force will increase to a maximum and then decrease along the positive direction of the Z-axis. The electromagnetic force increases continuously by approaching the electromagnet and reaches the maximum value at the end surface.

Based on the simulation of the single-coil electromagnet, simulation research on the 6 cascaded coil electromagnets has been carried out, which will lay the foundation for developing the control strategy of force generation. For the 6 coil electromagnets (EM1-EM6 in Fig. 9), we use $I1$-$I6$ to denote their driving currents respectively. When $I1$=$I2$=$I3$=1.6A and $I4$=$I5$=$I6$=-1.6A, i.e., the left three electromagnets and the right three electromagnets are driven with the same magnitude current with opposite directions, the magnetic field and the magnetic flux density distribution of the system are shown in Fig. 10 and Fig. 11. Specifically, the magnetic flux density of the point $P1$ (Z=112.5 mm, Y=0mm), i.e., the central point on the Z-axis between EM3 and EM4, is zero, because the electromagnetic fields generated by the driving currents on the left and right sides of this point are superimposed and lead to a perfect counteraction. According to simulation results, the magnetic flux density reaches a maximum value of 41.31 mT at Z=112.5 mm and Y=110 mm ($P2$), i.e. the inner edge side of the electromagnets.

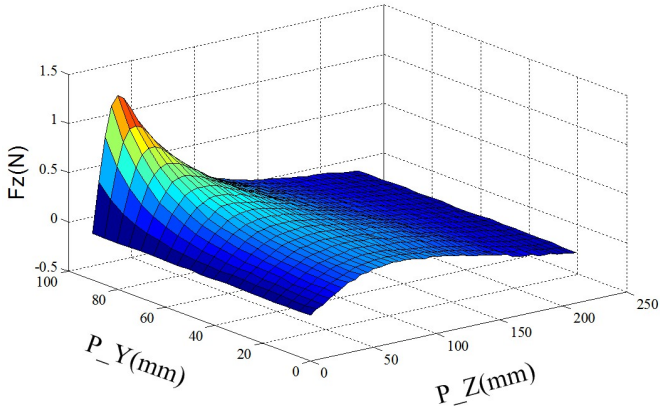

Fig. 8. Simulation of electromagnetic force (I=1.6A)

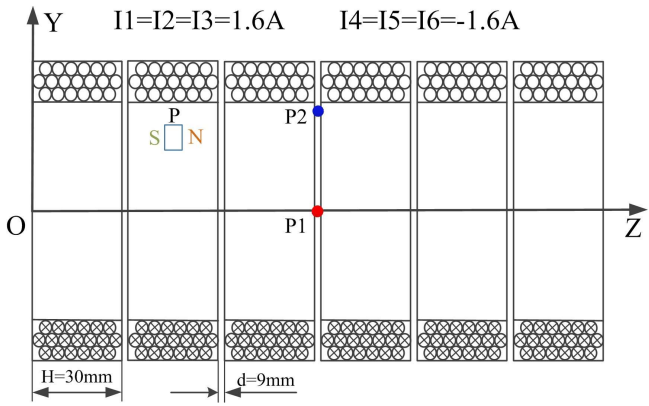

Fig. 9. The model diagram of cascaded electromagnets.

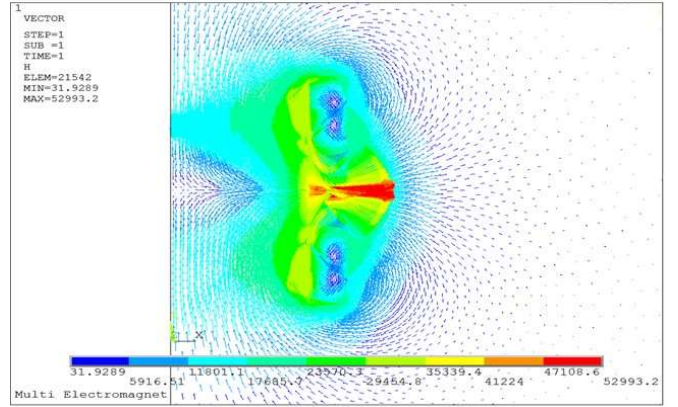

Fig. 10. Vector distribution of the electromagnetic field for six cascaded electromagnets

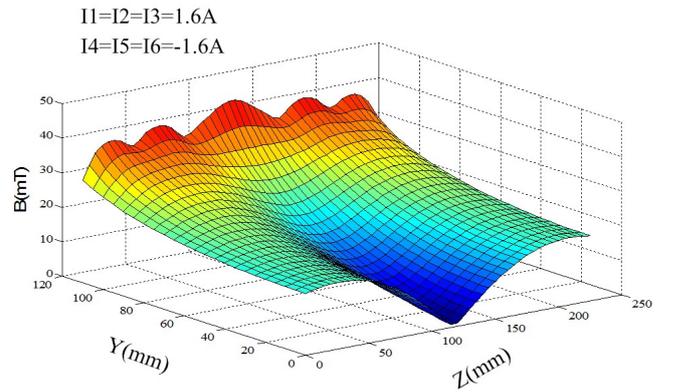

Fig. 11. Distribution of magnetic flux density for six cascaded electromagnets.

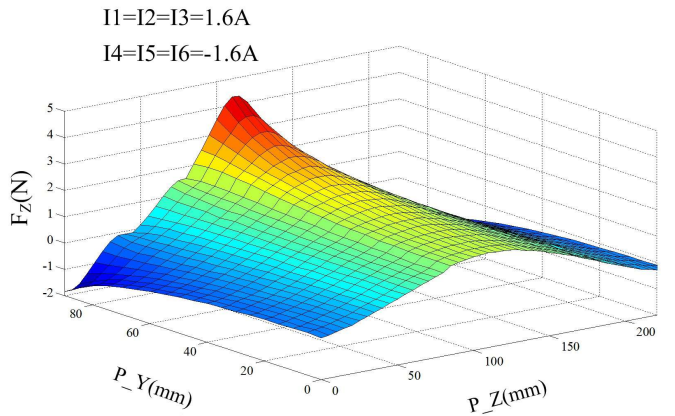

Fig. 12. Simulation of electromagnetic force for six cascaded electromagnets.

With the same driving currents, the simulation result of repulsive electromagnetic force in the Z-axis exerted on the permanent magnet (with the given position and alignment in Fig. 9) is shown in Fig. 12. Given $P\_Y$ of permanent magnet remains unchanged, the electromagnetic force will reach the maximum value where the magnetic fields change most rapidly, i.e., between EM3 and EM4 in the configuration of Fig. 9. The electromagnetic force reaches a maximum value of 4.31 N at the position of ($P\_Y$=110 mm, $P\_Z$=112.5 mm).

### B. Research on strategy for electromagnetic force control

The most critical challenge of the proposed haptic system is to determine the driving currents for 6 coils for any applicable magnetic force in the Z-axis direction and any arbitrary position of the permanent magnet. To illustrate our control strategy for generating the electromagnetic force, an expression is established as equation 1.

$$F_z = f(x, y, z, I_1, I_2, ..., I_6) = \sum_{I=1}^{6} g(x_i, y_i, z_i) D_i I_m \quad (1)$$

In equation 1, $F_z$ means the electromagnetic force imposing on the permanent magnet along the Z-axis direction; $I_1$-$I_6$ represent the driving currents applied to the electromagnets $EM1$-$EM6$ respectively. $D_i$ is the duty cycle of the PWM signal of $I_i$ and $I_m$ is the maximum driving current, i.e., 1.6A in our system. $g(x_i, y_i, z_i)$ denotes the transformation function between the electromagnetic force and the driving current of the electromagnet Emi (i=1…6), where (x, y, z) represents the position of the permanent magnet.

As the driving current for each single-coil electromagnet gradually increases from 0 to 1.6A, the resulting force on the permanent magnet at a fixed position will increase to a maximum amount accordingly. During each haptic rendering update loop, the calculated interacting forces will be converted to the driving current $I_i$, i.e., PWM duty cycle value, for each coil, through the real-time interpolation with the offline simulation data obtained previously.

To generate a magnetic force on the permanent magnet at a certain position, there are usually several control methods in the proposed system, i.e., driving a different number of single-coil electromagnets. Since the six cascaded electromagnets can each be independently driven in our proposed system, not only the magnitude but also the shape of the generated magnetic field can be regulated. To simplify the control method, a minimum-energy-consumption based strategy is used, i.e., our control strategy is always generating the desirable electromagnetic force with a minimum number of single-coil electromagnets.

During each force-generating loop, the priority of each single-coil electromagnet is determined according to its distance to the permanent magnet in the Z direction (Fig. 9), where the smaller the distance, the higher the coil's priority will be. Then the coil with the highest priority is chosen to generate the desired magnetic force by regulating its driving. If the desired force exceeds the maximum force generation capacity of the currently selected coils, the coil with the next lower priority will be used. The procedure will continue until the desired force is generated or all 6 single-coil electromagnets have been chosen and actuated with proper currents.

## V. EXPERIMENTS AND RESULT ANALYSIS

To verify our proposed contactless 1D magnetic haptic display system, several experiments were carried out. The experimental environment of our developed 1D magnetic haptic display is shown in Fig. 13, where a Microsoft HoloLens is used for visual rendering, and an LCD monitor is used to show real-time content of the HoloLens.

The virtual environment is designed in Unity 3D, using Microsoft Developer Toolkit HoloToolkit. And a Visual Studio 2013 WPF project is developed for implementing Kinect-based position tracking, where an Intel NUC is used as the calculating unit with an update rate of 27Hz. Through the established socket wireless communication based on the TCP protocol, the obtained position of the permanent magnet is sent from the NUC to the HoloLens, then the HoloLens implements the collision detection algorithm, and the calculated force is sent back to the NUC. At the same time, the visual scene is refreshed in the HoloLens in real-time. Based on the offline simulation data, the force is decoupled into the duty cycle values of six electromagnets, and sent to the SA306 based electromagnet control module through the serial communication, thereby generating 12 sets of PWM signals to drive six coils and form the controllable electromagnetic field.

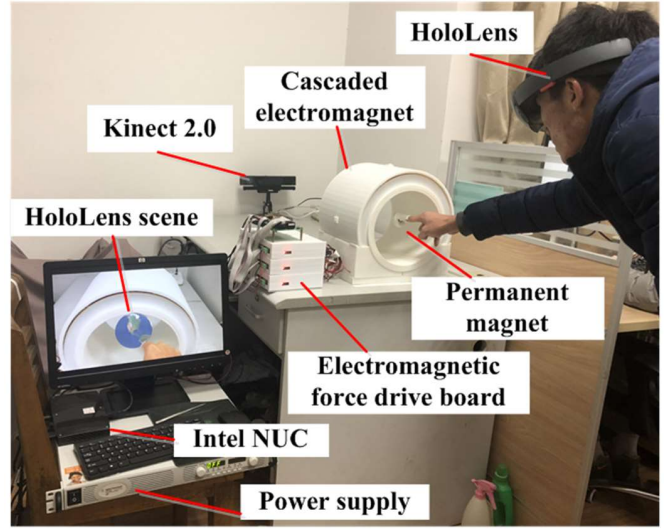

Fig. 13. 1D magnetic haptic system.

Five healthy volunteers (three males and two females), aging from 23 to 27 with a typical sense of touch and vision, participated in the experiments. Before each experiment, the main purpose and procedure of each experiment were introduced to the subjects. Then, each subject was given the warm-up time to get familiar with the system, which reduces the impact of possible uncertain factors to maintain the objectivity of the experimental results. The experiment received the approval of the Institutional Review Board of Nanjing University of Aeronautics and Astronautics. The participants gave their informed consent before the start of the experiments.

### A. Object recognition experiment

In this experiment, five subjects were asked to perceive and identify different objects without any visual feedback. As shown in Fig. 14, three simple 3D virtual objects were designed, including a sphere (with the diameter of 100mm), a cube (with the edge of 100mm), and a cylinder (with the diameter of 100mm and the length of 100mm). The permanent magnet can receive electromagnetic force on the surface and inside of the virtual objects. Before the experiment was carried out, the three virtual objects were shown one after another respectively to subjects with the HoloLens glass and the proposed 1D magnetic haptic system. For each virtual object, subjects were given time to perceive and feel it arbitrarily with the sense of touch. During the experiment, the HoloLens glass was turned off. Consequently, each subject was asked to identify a total of 30 virtual objects. Recognition results were recorded for each trial and subjects were not informed whether their answers were correct. Table II shows the recognition success rate of three virtual objects.

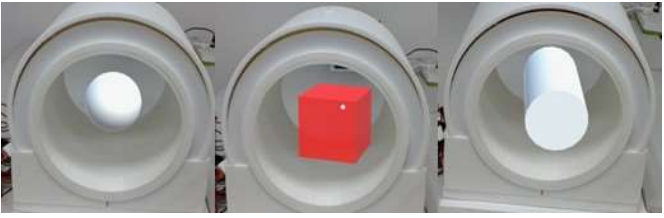

Fig. 14. Virtual objects including sphere, cube, and cylinder.

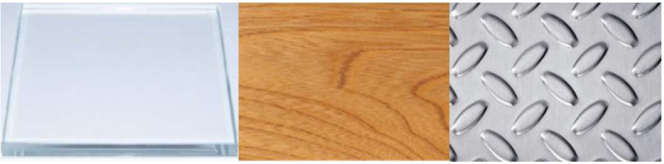

Fig. 15. Three virtual friction levels (Left: glass, Middle: wood, Right: steel)

TABLE II
SUCCESS RATE OF RECOGNITION VIRTUAL OBJECTS (%)

| Subjects | S1 | S2 | S3 | S4 | S5 | Average |
|---|---|---|---|---|---|---|
| Sphere | 100 | 100 | 100 | 100 | 100 | 100 |
| Cube | 90 | 80 | 90 | 90 | 90 | 88 |
| Cylinder | 90 | 90 | 80 | 90 | 100 | 90 |
| Subjects average identification rate | 93.33 | 90 | 90 | 93.33 | 96.67 | 92.67 |

As shown in Table II, a higher recognition rate is achieved for continuously curved surfaces, such as sphere and cylinder objects. However, the recognition success rate for the cube is lower. The reason for this might be the edge perception of the cube is not fine and sensitive enough for the subjects.

### B. Surface identification experiment

To verify the effectiveness of the system's tactile perception, a surface tactile identification experiment was further carried out. Three different levels of surface friction, L-1, L-2, and L-3, have been designed, which aim to simulate the surfaces of the glass, the wood, and the steel in the real world, respectively (Fig. 15). We use SFS (shape from shading) based method [48] to generate the haptic model for these three surfaces.

The subjects were asked to feel, touch, and then identify the currently rendered surfaces. Each of the three surfaces was randomly perceived for 10 times, and the identification time is not limited during the experiment. The obtained identification success rates for five subjects are shown in Table III.

It can be shown from Table III that the contactless magnetic haptic system proposed in this paper can successfully render the surfaces with different frictions. At the same time, it can be seen that a distinct friction mode (like the steel diamond plate in the experiment) will contribute to a higher identification success rate.

TABLE III
SUCCESS RATE OF SURFACE IDENTIFICATION (%)

| Subjects | S1 | S2 | S3 | S4 | S5 | Average |
|---|---|---|---|---|---|---|
| Level-1 glass | 100 | 100 | 100 | 90 | 100 | 98 |
| Level-2 wood | 100 | 90 | 100 | 100 | 100 | 98 |
| Level-3 steel | 100 | 100 | 100 | 100 | 100 | 100 |

### C. User perception evaluation experiment

In order to further verify the effectiveness of the proposed magnetic haptic interface, an experiment for evaluating the haptic perception of the subjects for haptic-based interactions has been carried out. Wearing the Microsoft HoloLens device and our designed fingertip-mounted permanent magnet, subjects were required to touch virtual objects and rate the reality of their perception compared with their experience in the real world. The reality for haptic perception is divided into five levels (grade5-grade1), which indicates the perception is exactly the same as (grade5), almost the same as (grade4), a little different with (grade3), moderately different with (grade2), and quite different with (grade1) that in the real world, respectively.

The experiment result is shown in Table IV. The average score of 3.6 demonstrates that a relatively realistic haptic perception can be rendered with the proposed magnetic haptic device.

TABLE IV
RESULT OF PERCEPTION REALITY EVALUATION OF THE PROPOSED MAGNETIC HAPTIC INTERFACE.

| Subjects | S1 | S2 | S3 | S4 | S5 | Average |
|---|---|---|---|---|---|---|
| Rating score | 3 | 4 | 4 | 3 | 4 | 3.6 |

### D. Discussion

Due to the cascaded structure for the six identical hollow disk electromagnets in our proposed magnetic haptic device, a relatively large cylindrical workspace of $\varnothing 210mm * H 234mm$ is achieved. The maximum force of 0.39 N for driving a single electromagnet and 4.31N for driving all six electromagnets simultaneously can be generated. With the precomputed FEM-based simulation data reflecting the relationship between the magnetic force, fingertip permanent magnet position, and driving currents, a typical force update rate of 500Hz or 1000Hz can be obtained. However, due to large coil-based electromagnets being used for regulating magnetic forces, we found through experiments that the maximum response time for changing the driving current from 0 A to 1.6A for a single electromagnet was about 40ms. This current frame rate of 25Hz still suffices for many interactive scenarios for rendering virtual objects with a limited stiffness.

## VI. CONCLUSION AND FUTURE WORKS

A contactless electromagnetic force based haptic display method is developed and evaluated experimentally in this paper. This untethered 1D haptic device realizes a non-contact HCI method, where the human operator can interact with the virtual object intuitively and naturally with his/her bare finger. Taking a new approach, we have utilized cascaded electromagnets modules composed of six separate coils to generate controllable electromagnetic forces. The current of each coil can be controlled separately, which enables the control of the cascaded electromagnets in a more flexible and convenient manner. In addition, an optimal control strategy for generating electromagnetic force is proposed based on offline simulation data, which also realizes haptic feedback with a minimum number of electromagnet coils. The virtual object recognition experiment, virtual surface identification experiment, and the user perception evaluation experiment have been carried out, which verify the functions and validity of the proposed system. Due to its relatively large workspace and contactless interaction mode, the proposed haptic interface has potential applications in various fields, such as haptic-based teleoperation, virtual medical training, gaming or collaborative robotics.

The electromagnetic force-based haptic device is a promising method for natural and contactless haptic rendering, which provides the benefits of high mobility, fewer constraints, and intuitive interactions. However, there are still several challenges and difficulties in the design of electromagnetic haptic systems. Real-time and accurate fingertip position racking methods for use in the magnetic environment will no doubt improve the system performance of magnetic haptic inter- faces [49]. While improved forms of hand and finger gesture recognition [50] can further enhance the user experience. A more superior electromagnetic force control strategy could be proposed to obtain a shorter response time and less energy consumption. At the same time, based on this multi-modal system integrating vision and haptic feedback, more extensive experiments and applications for medical simulation or education are needed.

It will be of more usable and practical importance if magnetic forces in 3D can be realized. Furthermore, human hand-based multi-point haptic rendering will further improve the quality and efficiency of human-computer interactions and further extends the application fields of haptic rendering. Alternative locations of magnets on the hand might also be explored such as on the finger-nail [51], thumb or ring [52]. In HCI, the realization of such contactless haptic rendering across workspaces, could also afford new discreet interactions with invisible magnetic forces.

## ACKNOWLEDGMENT

The authors would like to gratefully acknowledge the contribution of reviewers' comments.

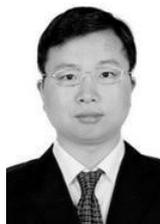

**Xiong Lu** received his BS, MS and Ph.D. degree from the Southeast University, Nanjing, China, in 2000, 2003 and 2008, respectively. He is currently an Associate Professor with the College of Automation Engineering, Nanjing University of Aeronautics and Astronautics. He has been a Visiting Scholar (2019-2020) at Computer Human Interaction (SACHI) Group, School of Computer Science, University of St Andrews, Scotland, United Kingdom. His current research interests include haptic rendering, tactile rendering and human computer interaction.

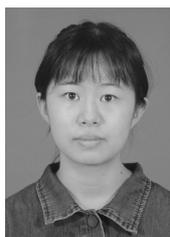

**Yuxing Yan** is currently pursuing her M.S. degree with the College of Automation Engineering, Nanjing University of Aeronautics and Astronautics. Her current research interests include haptic rendering and human computer interaction.

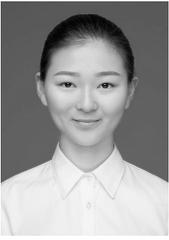

**Beibei Qi** has received her M.S. degree with the College of Automation Engineering, Nanjing University of Aeronautics and Astronautics. Her current research interests include haptic rendering and human computer interaction.

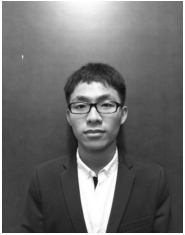

**Huang Qian** has received his M.S. degree with the College of Automation Engineering, Nanjing University of Aeronautics and Astronautics. His current research interests include haptic rendering and human computer interaction.

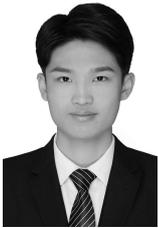

**Junbin Sun** has received his M.S. degree with the College of Automation Engineering, Nanjing University of Aeronautics and Astronautics. His current research interests include haptic rendering and human computer interaction.

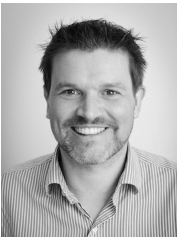

**Aaron Quigley** is a Professor and Head of School of Computer Science and Engineering, UNSW, in Sydney Australia. Aaron directs the BODi Lab and is the technical program chair for the ACM EICS 2022. He was the general co-chair for the ACM CHI Conference on Human Factors in Computing Systems in Yokohama Japan (online) in 2021. His research interests include discreet computing, global HCI, pervasive and ubiquitous computing and information visualisation.